\documentclass[final]{aipproc}

\layoutstyle{8x11double}

\begin{document}

\title{Towards the continuum limit of the lattice Landau gauge gluon propagator}


\author{O. Oliveira}{
  address={Centro de F\'{i}sica Computacional, Rua Larga, Universidade de Coimbra, 
P-3004-516 Coimbra, Portugal},altaddress={Departamento de F\'{\i}sica, Instituto Tecnol\'ogico de Aeron\'autica, 12.228-900, S\~ao Jos\'e dos Campos, SP, Brazil }
}

\author{P. J. Silva}{
  address={Centro de F\'{i}sica Computacional, Rua Larga, Universidade de Coimbra, P-3004-516 Coimbra, Portugal}
}

\begin{abstract}
The infrared behaviour of the lattice Landau gauge gluon propagator is discussed, combining results from simulations with different volumes and lattice spacings. In particular, the Cucchieri-Mendes bounds are computed and their implications for D(0) discussed.
\end{abstract}

\maketitle


\section{Introduction and motivation}

The link between the deep infrared behaviour of the gluon and ghost propagators and confinement,
has motivated a great effort on computing these quantities on the lattice. Besides
checking gluon confinement criteria, another important goal is to compare 
recent solutions of the Dyson-Schwinger equations with lattice results. In particular, the 
scaling solution \cite{fischer06} predicts a vanishing gluon propagator and a 
divergent ghost propagator at zero momentum. This solution complies with 
Gribov-Zwanziger \cite{zwanziger} and Kugo-Ojima \cite{kugo} confinement 
criteria. 
On the other hand, the decoupling solution \cite{papava} claims 
that a finite and non-vanishing zero momentum gluon propagator and a tree level like ghost propagator. The value
of the zero momentum gluon propagator is connected with a dynamical generated gluon mass.

In this paper we report on our current results for the Cucchieri-Mendes 
bounds in SU(3) lattice gauge theory.

\section{Cucchieri-Mendes bounds}

The Cucchieri-Mendes bounds \cite{Cuc0712} provide upper and lower bounds for the zero momentum gluon propagator of lattice Yang-Mills theories in terms of the average value of the gluon field. In particular, they relate the gluon propagator at zero momentum $D(0)$ with
\begin{equation}
   M(0) ~ = ~ \frac{1}{d \left( N^2_c - 1 \right)} \sum_{\mu, a} \left| A^a_\mu (0) \right| \, ,
\end{equation}   
where $d$ is the number of space-time dimensions, and $N_c$ the number of colors. In the above equation, $A^a_\mu (0)$ is the $a$ color component of the gluon field at zero momentum, defined by
\begin{equation}
  A^a_\mu (0) ~ = ~ \frac{1}{V} \sum_x A^a_\mu (x)
\end{equation}  
where $A^a_\mu (x)$ is the $a$ color component of the gluon field in the real space. 
 $D(0)$ is related with $M(0)$ by
\begin{equation}
  \langle M(0) \rangle^2 ~ \le ~ \frac{D(0)}{V} ~ \le N_d \left(N^2_c - 1\right) \langle M(0)^2 \rangle \, .
  \label{bounds}
\end{equation}  
In the last equation $\langle ~ \rangle$ means Monte Carlo average over gauge configurations. For convenience we will use the definition $N_{cd}=N_{d}(N_{c}^2-1)$. The
bounds in equation (\ref{bounds}) are a direct result of the Monte Carlo approach. 
The interest on these bounds comes from allowing a scaling analysis which can help understanding
the finite volume behaviour of $D(0)$: assuming  that each of the terms in inequality (\ref{bounds}) scales with the volume according to $ A / V^\alpha$, the simplest possibility 
and the one considered in \cite{Cuc0712}, an $\alpha > 1$  for $ \langle M(0)^2 \rangle$ clearly indicates that $D(0) \rightarrow 0$ as the infinite volume is approached. In this sense, this scaling analysis allows to investigate the behaviour of $D(0)$ in the infinite volume limit. 

 For the SU(2) Yang-Mills theory \cite{Cuc0712},  the results
show a $D(0)=0$ for the two dimensional theory, but a $D(0) \ne 0$ for three and four dimensional formulations.

\section{Results for SU(3) gauge theory}

We have studied the Cucchieri-Mendes bounds within SU(3) lattice gauge theory for three values of the gauge coupling: $\beta=6.0$ \cite{bounds_su3, lat09}, $\beta=5.7$ \cite{lat09}, and $\beta=6.2$.

\subsection{Scaling analysis for $\beta=6.0$}

\begin{figure}[!t]
  \includegraphics[width=0.45\textwidth]{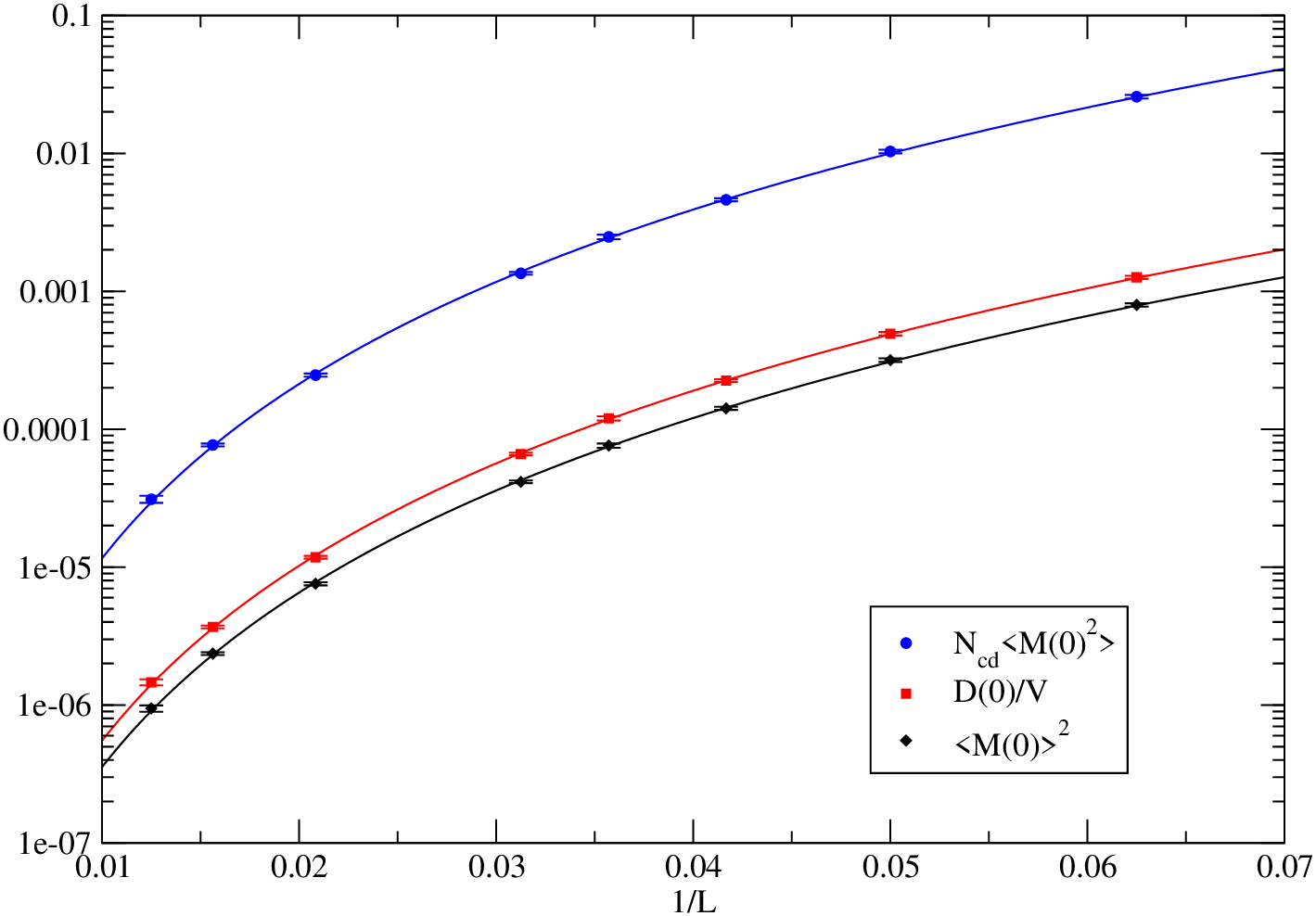}
  \caption{Cucchieri-Mendes bounds for $\beta=6.0$.}
\label{bounds60}
\end{figure}

In table \ref{setup60} we present the lattice setup for $\beta=6.0$, pointing out the differences to \cite{bounds_su3, lat09}.

\begin{table}
\begin{tabular}{c@{\hspace{0.2cm}}c@{\hspace{0.2cm}}c@{\hspace{0.2cm}}c@{\hspace{0.2cm}}c@{\hspace{0.2cm}}c@{\hspace{0.2cm}}c@{\hspace{0.2cm}}c@{\hspace{0.2cm}}c}
\hline
$L^4$  &  $16^4$ & $20^4$ & $24^4$ & $28^4$ & $32^4$  & $48^4$ & $64^4$ & $80^4$ \\
L(fm)  &   1.63  & 2.03   & 2.44   & 2.84  & 3.25    &  4.88  & 6.50 & 8.13 \\
\# conf. &   52    & 72     &  60\tablenote{new ensemble}& 56    & 126     &  104   & 120 & 50\tablenote{new statistics} \\
\hline
\end{tabular}
\caption{Lattice setup for $\beta=6.0$. The lattice spacing is $a=0.1016(25)$fm.}
\label{setup60}
\end{table}

Figure \ref{bounds60} shows the results for the bounds, together with the fits 
to $\omega/V^{\alpha}$. Assuming this simple scaling behaviour, our results for 
the exponent $\alpha$ support $D(0)=0$ -- see table \ref{fits.ppl.b60}. However, 
when one assumes a scaling behaviour like $C/V+\omega V^{-\alpha}$, the results 
support $D(0)\neq0$ -- see table \ref{fits.const.b60}. In this sense, a finite and 
non-vanishing value for $D(0)$ in the infinite volume is not excluded.

\begin{table}
\begin{tabular}{rrrr}
\hline
                           &  $\omega$ & $\alpha$   & $\chi^{2}_{\nu}$ \\
\hline
  $\langle M(0) \rangle$   &  9.53(36)  & 0.5255(26)  & 0.80  \\
  $D(0)/V$                 & $149\pm10$    & 1.0542(49)  & 0.63 \\
 $N_{cd}\langle M(0)^2 \rangle$ & $2927\pm221$  & 1.0504(54)  & 0.83  \\
 \hline
\end{tabular}
\caption{Fits to $\omega/V^{\alpha}$ using lattice data at $\beta=6.0$.}
\label{fits.ppl.b60}
\end{table}

\begin{table}
\begin{tabular}{rrrrr}
\hline
                                   &  $\omega/1000$   &  $\alpha$    &  $C/100$     & $\chi^{2}_{\nu}$ \\
\hline
  $\langle M(0) \rangle^2$              & $0.23(24)$ & $1.22(11)$   &   $0.337(50)$   &  0.47 \\
  $D(0)/V$                              & $0.27(23)$  & $1.19(10)$  &   $0.49(11)$     & 0.42  \\
  $N_{cd} \langle M(0)^2 \rangle$  &  $7.1\pm7.3$ & $1.22(11)$  &  $11.0\pm1.7$   & 0.55  \\
 \hline
\end{tabular}
\caption{Fits to $C/V+\omega V^{-\alpha}$ using lattice data at $\beta=6.0$. }
\label{fits.const.b60}
\end{table}

Concerning the fits to $\omega/V^{\alpha}$,  the reasons for the differences in the values of $\alpha$ reported 
here and in \cite{Cuc0712} -- and therefore on the behaviour of $D(0)$ in the infinite volume limit -- are not
clear. The simulations use different gauge groups. Although there it is generally believed that the SU(2) and
SU(3) propagators are equivalent for momenta above 1 GeV \cite{su23ptbr, su23aust}, a recent direct comparison for smaller momenta has shown a measurable difference in the infrared region \cite{trento}.

Moreover, the physical volumes used in \cite{Cuc0712} are much larger -- up to (27fm)$^4$ -- than the ones used here -- up to (8fm)$^4$. However, the reader should be aware that in the SU(2) case the lattice spacing used is about twice the lattice spacing considered here.

\section{Lattice spacing effects in the gluon propagator}

In order to disentangle possible lattice effects due to the use of a different lattice spacing, we carried out simulations at $\beta=5.7$ and $\beta=6.2$. The lattice setup is shown in tables \ref{setup57} and \ref{setup62} respectively.

\hspace*{0.5cm}

\begin{table}[!h]
\begin{tabular}{c@{\hspace{0.2cm}}c@{\hspace{0.2cm}}c@{\hspace{0.2cm}}c@{\hspace{0.2cm}}c@{\hspace{0.2cm}}c@{\hspace{0.2cm}}c@{\hspace{0.2cm}}c}
\hline
$L^4$     &  $8^4$  & $10^4$ & $14^4$ & $18^4$ & $26^4$  & $36^4$ & $44^4$ \\
L(fm)     &   1.47  & 1.84   & 2.57   & 3.31  & 4.78    &  6.62  & 8.09  \\
\# conf.  &   56    & 149    &  149   & 149   & 132     &  100   &   55$^{*}$   \\
\hline
\end{tabular}
\caption{Lattice setup for $\beta=5.7$. The lattice spacing is $a=0.1838(11)$fm.}
\label{setup57}
\end{table}

\hspace*{0.5cm}

\begin{table}[!h]
\begin{tabular}{cccccccc}
\hline
$L^4$     &  $24^4$ & $32^4$ & $48^4$ & $64^4$ & $80^4$   \\
L(fm)     &  1.74   & 2.32   &  3.49  & 4.65  &   5.81    \\
\# conf.  &  51     &   56   &  87    & 99   &   15     \\
\hline
\end{tabular}
\caption{Lattice setup for $\beta=6.2$. The lattice spacing is $a=0.07261(85)$fm.}
\label{setup62}
\end{table}

\hspace*{0.5cm}

\begin{figure}[t]
\includegraphics[width=0.45\textwidth]{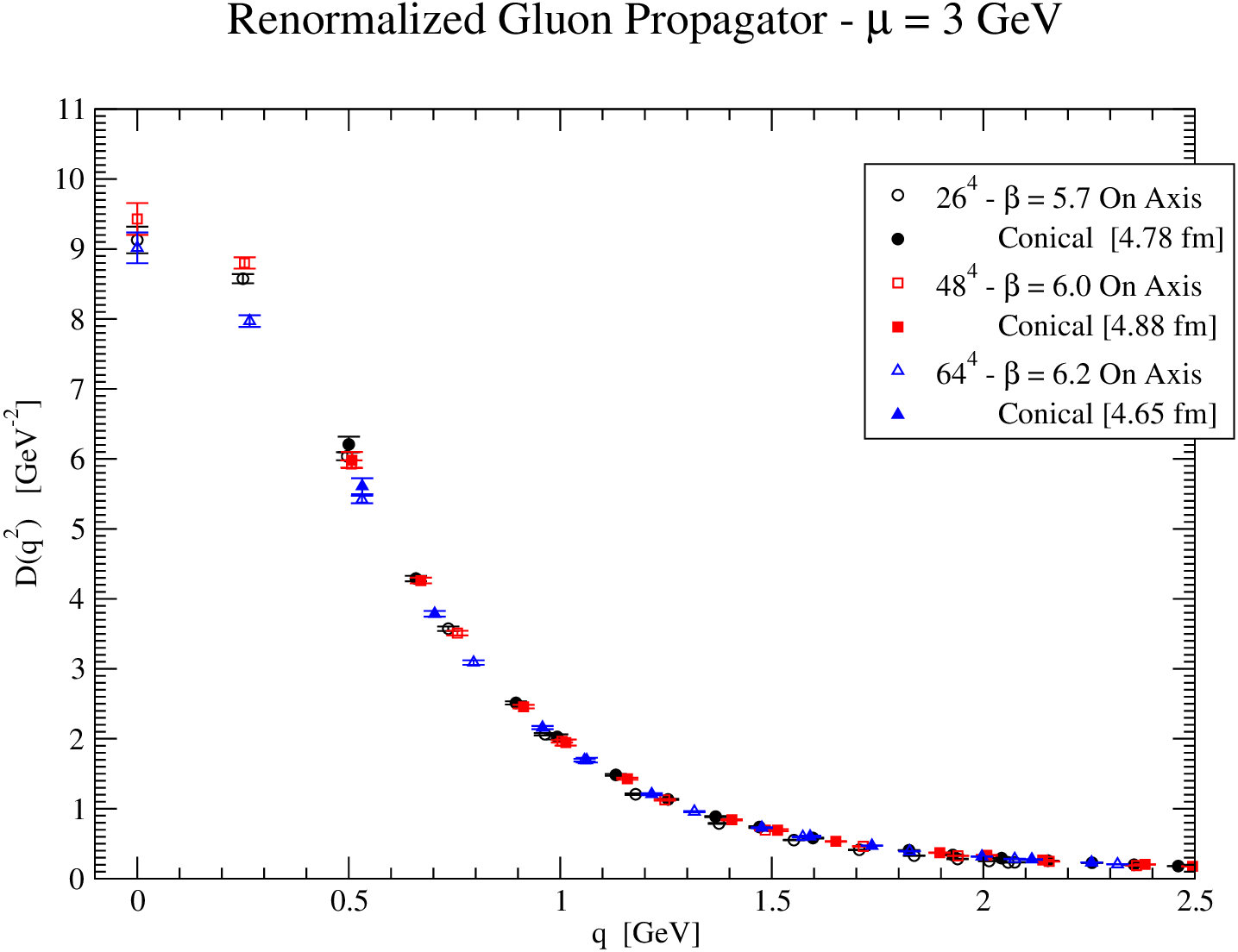}
\caption{Comparing the gluon propagator computed \mbox{using} different lattice spacings at the same physical volume $V\sim(4.8fm)^4$.}
\label{gluvol48}
\end{figure}

\begin{figure}[t]
\includegraphics[width=0.45\textwidth]{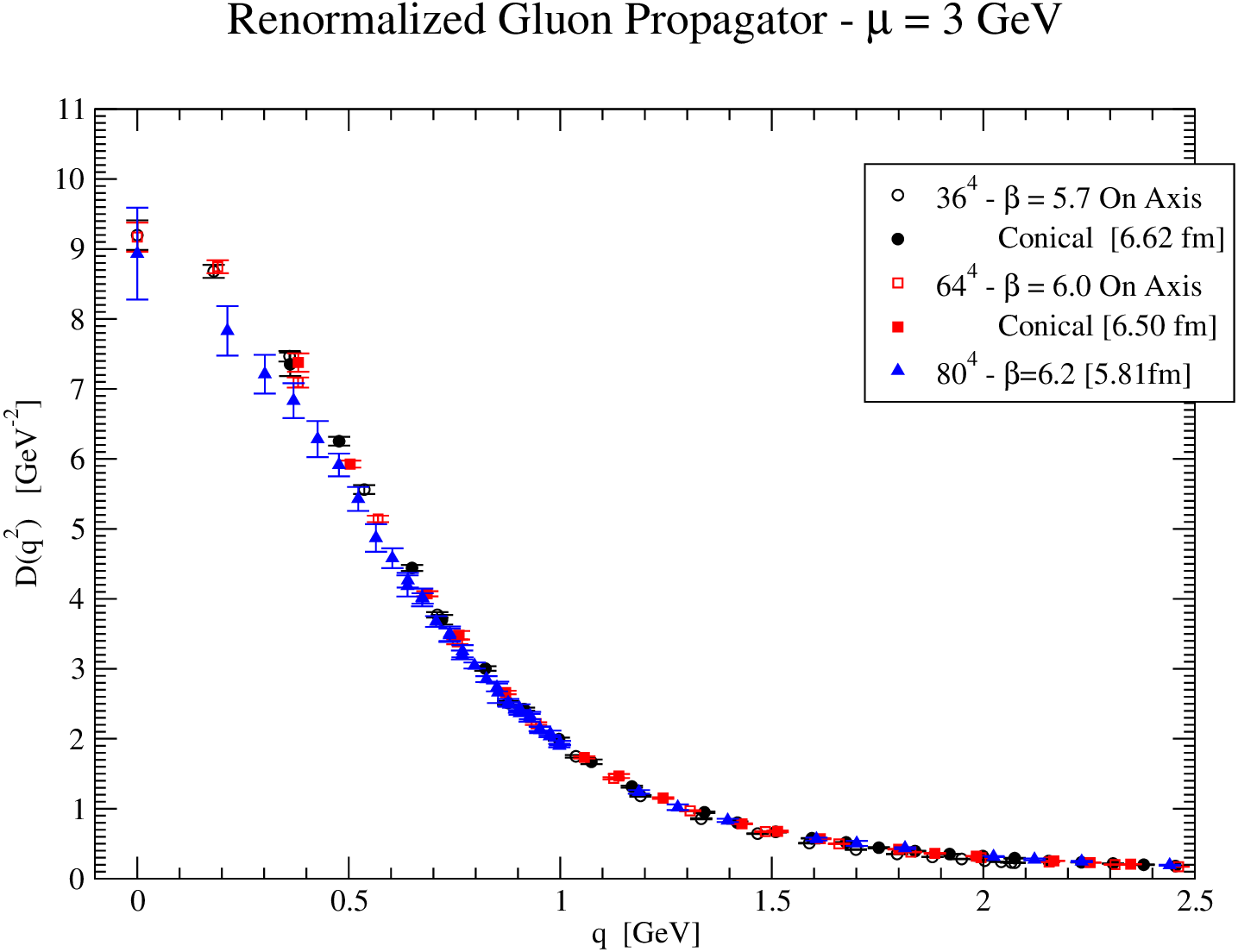}
\caption{Comparing the gluon propagator computed \mbox{using} different  lattice spacings at the same physical volume $V\sim(6.5fm)^4$.}
\label{gluvol65}
\end{figure}

Some differences have been seen between the gluon propagator computed at 
different lattice spacings for similar physical volumes. An example 
can be seen in figures \ref{gluvol48} and \ref{gluvol65}, where the 
infrared $\beta=6.2$ data does not agree with data from $\beta=5.7$ 
and $6.0$ simulations.
These differences deserve further investigations to clarify any possible 
effects due to finite lattice spacing.

\subsection{Scaling analysis for $\beta=5.7 $ and $\beta=6.2$}

In what concerns the fits to $\omega/V^{\alpha}$, the analysis of the data coming from both sets still supports a vanishing $D(0)$ in the infinite volume limit -- see tables \ref{fits.ppl.b57} and \ref{fits.ppl.b62}. 

Similarly to the case studied before, the lattice data is also well described by the functional form $C/V+\omega V^{-\alpha}$ -- see tables \ref{fits.const.b57} and \ref{fits.const.b62}. Although the $\beta=5.7$ case supports $D(0)\neq0$, for $\beta=6.2$ the statistical errors do not allow to take any conclusion. In fact, although $C=0$ within statistical errors, we also get $\alpha=1$. For this case, it is worth an increase of statistics.

\begin{table}[t]
\begin{tabular}{rrrr}
\hline
                           &  $\omega$  & $\alpha$    & $\chi^{2}_{\nu}$ \\
\hline
  $\langle M(0) \rangle$           &  4.63(12)   & 0.5244(23)    & 1.92 \\
  $D(0)/V$                         & $32.8\pm1.6$  & 1.0466(42)   & 1.14 \\
 $N_{cd}  \langle M(0)^2 \rangle$ & $696\pm37$  & 1.0488(47)  & 1.72 \\
 \hline
\end{tabular}
\caption{Fits to $\omega V^{-\alpha}$ using lattice data at $\beta=5.7$. In order to keep $\chi^{2}_{\nu}<2$, the $26^4$ lattice data has been excluded.}
\label{fits.ppl.b57}
\end{table}

\vspace*{1cm}

\begin{table}[b]
\begin{tabular}{rrrr}
\hline
                           &  $\omega$/100  & $\alpha$    & $\chi^{2}_{\nu}$ \\
\hline
  $\langle M(0) \rangle$           &  $0.163(11)$   & 0.5374(47)    & 0.08 \\
  $D(0)/V$                         & $3.66(46)$  & 1.0659(84)   & 0.47 \\
 $N_{cd} \langle M(0)^2 \rangle$ & $8.4\pm1.2$  & 1.0725(94)  & 0.13  \\
 \hline
\end{tabular}
\caption{Fits to $\omega V^{-\alpha}$ using lattice data at $\beta=6.2$. Data for $M(0)$ does not include $48^4$.}
\label{fits.ppl.b62}
\end{table}

\begin{table}[t]
\begin{tabular}{r@{\hspace{0.3cm}}r@{\hspace{0.3cm}}r@{\hspace{0.3cm}}r@{\hspace{0.3cm}}r}
\hline
                                   &  $\omega/100$   &  $\alpha$    &  $C/100$     & $\chi^{2}_{\nu}$ \\
\hline
  $\langle M(0) \rangle^2$         & $0.27(15)$ & $1.186(90)$  & $0.088(15)$   &  1.80 \\
  $D(0)/V$                         & $0.301(93)$  & $1.122(90)$ & $0.116(53)$  & 1.28  \\
  $N_{cd} \langle M(0)^2 \rangle$   & $8.2\pm4.2$ & $1.172(91)$  & $2.78(58)$   & 1.69  \\
 \hline
\end{tabular}
\caption{Fits to $C/V +\omega V^{-\alpha}$ using lattice data at $\beta=5.7$. In order to keep $\chi^{2}_{\nu}<2$, the $26^4$ lattice data has been excluded.}
\label{fits.const.b57}
\end{table}

\vspace*{-1.5cm}
\begin{table}[b]
\begin{tabular}{rrrrr}
\hline
                              &  $\omega/1000$   &  $\alpha$    &  $C/100$     & $\chi^{2}_{\nu}$ \\
\hline
  $\langle M(0) \rangle^2$       & $0.34(66)$   & $1.13(29)$ & $0.4\pm1.2$     &  0.13 \\
  $D(0)/V$                       & $0.366(47)$  & $1.07(29)$ & $0.04\pm5.6$      & 0.95   \\
  $N_{cd}\langle M(0)^2 \rangle$  & $8.6\pm6.7$  & $1.08(28)$ & $4\pm85$   & 0.25  \\
 \hline
\end{tabular}
\caption{Fits to $C/V+\omega V^{-\alpha}$ using lattice data at $\beta=6.2$. Data for $M(0)$ does not include $48^4$.}
\label{fits.const.b62}
\end{table}

\section{Conclusions}

We have studied the scaling behaviour of Cucchieri-Mendes bounds using ensembles generated at several lattice spacings. Fits of the data to a pure power law in the volume strongly support $D(0)=0$, but the use of other ansatze do not allow to take definitive conclusions.


\begin{theacknowledgments}
The authors acknowledge financial support from FCT under contracts PTDC/FIS/100968/2008, CERN/FP/109327/2009 and (P.J.S.) grant SFRH/BPD/40998/2007. O.O. acknowledges financial support from FAPESP.
\vspace{1cm}
\end{theacknowledgments}

\end{document}